\documentclass[usenatbib]{mn2e}
\usepackage{graphicx}

\title{Rotational evolution of the Crab pulsar in the wind braking model}

\author[Kou \& Tong]
    {F. F. Kou and H. Tong
\\
    Xinjiang Astronomical Observatory, Chinese Academy of Sciences, Urumqi, Xinjiang 830011,
    China; {\it tonghao@xao.ac.cn}
\\}

\begin{document}

\date{2015.3 v3}
\pagerange{\pageref{firstpage}--\pageref{lastpage}} \pubyear{2015}
\maketitle

\label{firstpage}
\begin{abstract}
The pulsar wind model is updated by considering the effect of
particle density and pulsar death. It can describe both the short
term and long term rotational evolution of pulsars consistently.
It is applied to model the rotational evolution of the Crab
pulsar. The pulsar is spun down by a combination of magnetic
dipole radiation and particle wind. The parameters of the Crab
pulsar, including magnetic field, inclination angle, and particle
density are calculated. The primary particle density in
acceleration region is about $10^{3}$ times the Goldreich-Julian
charge density. The lower braking index between glitches is due to
a larger outflowing particle density. This may be glitch induced
magnetospheric activities in normal pulsars. Evolution of braking
index and the Crab pulsar in $P-\dot{P}$ diagram are calculated.
The Crab pulsar will evolve from magnetic dipole radiation
dominated case towards particle wind dominated case. Considering
the effect of pulsar ``death'', the Crab pulsar (and other normal
pulsars) will not evolve to the cluster of magnetars but downwards
to the death valley. Different acceleration models are also
considered. Applications to other sources are also discussed,
including pulsars with braking index measured, and the magnetar
population.
\end{abstract}


\section{Introduction}           

The Crab pulsar (PSR B0531+21) is a young radio pulsar with a spin
frequency $\nu=30.2 \, \rm Hz$ and frequency derivative
$\dot{\nu}=-3.86 \times10^{-10} \, \rm Hz/s$ (Lyne et al. 1993).
Its  characteristic magnetic field  is about $7.6 \times10^{12} \,
\rm G$ at the magnetic poles\footnote{Assuming all the rotational
energy loss is due to magnetic dipole radiation in vaccum, $B_{\rm
c}=6.4 \times 10^{19} \sqrt{P \dot{P}} \rm \, G$}. The Crab pulsar
has been monitored continuously for decades years by various
telescopes since it was discovered in 1968 (Lyne et al. 1993; Lyne
et al. 2015; Wang et al. 2012). Its braking index is $n=2.51 \pm
0.01$ (Lyne et al. 1993). Different braking index values $2.45$,
$2.57$ and $2.3$ are also reported between glitches (Wang et al.
2012; Lyne et al. 2015). Observational details are given in Table
\ref{parameters}.

Long-term observations have found that almost all pulsars (except
accreting X-ray pulsars in binary systems) are spinning down. The
spin-down behavior can be described by a power law (Lyne et al.
1993):
\begin{equation}\label{nudotpowerlaw}
 \dot{\nu}=-k\nu^{n},
\end{equation}
here, $k$ is usually taken as a constant and $n$ is the braking
index. The braking index and second braking index are defined
respectively (Livingstone et al. 2005):
\begin{equation}
n=\frac{\nu \ddot{\nu}}{\dot{\nu}^{2}},
\label{defn}
\end{equation}
\begin{equation}
m=\frac{\nu^{2}\stackrel{...}{\nu}}{\dot{\nu}^{3}},
\label{defm}
\end{equation}
where $\nu$ is the pulsar spin frequency,  $\dot{\nu}$,
$\ddot{\nu}$ and $\stackrel{...}{\nu}$ are the first, second and
third frequency derivative, respectively. The spinning down of
pulsars is usually assumed to be braked by the magnetic dipole
radiation (Shapiro \& Teukolsky 1983). In the magnetic dipole
radiation model, a neutron star rotates uniformly in vacuo at a
frequency $\nu$ and possesses a magnetic moment $\mu$. The
corresponding slowdown rate is:
\begin{equation}
\dot{\nu}=-\frac{8 \pi^{2}\mu^{2}}{3 I c^{3}}\nu^{3} \sin^{2}
\alpha,
\end{equation}
where $\mu=1/2 B R^{3}$ is the magnetic dipole moment ($B$ is the
polar magnetic field and $R$ is the radius of neutron star),
$I=10^{45}\, \rm {g \cdot cm^{2}}$ is the moment of inertia, $c$
is the speed of light, and $\alpha$ is the angle between the
rotational axis and the magnetic axis (i.e., the inclination
angle). The spin down behavior of pulsar in this model can be
described as $\dot{\nu}\propto\nu^{3}$. The braking index is
exactly three if $\mu, \ I$, and $\alpha$ are constant. To date,
only eight pulsars, including the Crab pulsar, have measured the
meaningful braking indices for they are young and own relatively
larger $\dot{\nu}$ (Espinoza et al. 2011; Lyne et al. 2015). Their
braking indices are all smaller than three. It means that there
are other physical processes needed to slow down the pulsar.

Many mechanisms have been proposed in order to explain the braking
index observations, e.g., the pulsar wind model (Xu \& Qiao 2001;
Wu et al. 2003; Contopoulos \& Spitkovsky 2006; Yue et al. 2007),
a changing magnetic field strength (Lin et al. 2004; Chen \& Li
2006; Espinoza et al. 2011), a changing inclination angle (Lyne et
al. 2013), and additional torques due to accretion (Liu et al.
2014). However, these models are more or less not consistent with
observations or can not simulate the long-term evolution of
pulsars. And these models can not explain the different braking
indices detected between glitches (Wang et al. 2012; Lyne et al.
2015). Furthermore, the effect of pulsar death (death line,
Ruderman \& Sutherland 1975; or  death valley, Chen \& Ruderman
1993; Zhang et al. 2000) should be considered in modeling the long
term rotational evolution of the pulsar (Contopoulos \& Spitkovsky
2006).

The pulsar wind model considers both the pulsar spin-down and the
particle acceleration in the magnetosphere (Xu \& Qiao 2001; Wu et
al. 2003). In this paper, an updated pulsar wind model is built
based on previous researches (Xu \& Qiao 2001; Yue et al. 2007;
Contopoulos \& Spitkovsky 2006; Li et al. 2012). It includes: (1)
Both magnetic dipole radiation and particle outflow, and their
dependence on the inclination angle; (2) The particle outflow
depends on the specific acceleration model; (3) The primary
particle density may be much larger than the Goldreich-Julian
charge density; (4) The effect of pulsar death is considered in
modeling the rotational evolution of the pulsar. This model can
calculate both the short term and long term evolution of pulsars.
It is applied to the Crab pulsar which has the most detailed
timing observations. Possible applications to other sources are
also illustrated.

The pulsar wind model and model calculations of the Crab pulsar are presented in section 2.
Discussions and conclusions are given in section 3 and section 4, respectively.

\begin{table}
\label{parameters}
\begin{center}
\caption{Observations of the Crab pulsar.}
\begin{tabular}{ll}
\hline \hline
Pulsar parameter & Values\\
\hline
Epoch(MJD) & $40000.0$\\
 $\nu(\rm Hz)$ & $30.225437 ^{a}$\\
 $\dot{\nu}(10^{-10} \rm Hz/s)$ & $-3.86228  ^{a}$ \\
$\ddot{\nu}(10^{-20} \rm Hz/s^{2})$ &  $1.2426 ^{a}$\\
$\stackrel{...}{\nu}(10^{-30} \rm Hz/s^{3})$ &  $-0.64 ^{a}$\\
Braking index  $n$ & $2.51 \pm 0.01^{a}$\\
Second braking index $m$ & $10.15^{a}$ \\
Inclination angle $\alpha(^\circ)$ & $(45 \sim 70)^{b}$\\
Age & $915 \, \rm yr \, (from \, 1054 \, to \, 1969)$\\
 \hline
\end{tabular}
\flushleft
Notes:
(a): Observed spin frequency and its derivatives are
parameters in the Taylor expansion
$\nu=\nu_0+\dot{\nu_0}(t-t_0)+\frac{1}{2}\ddot{\nu_0}(t-t_0)^{2}+\frac{1}{6}\stackrel{...}{\nu_0}(t-t_0)^{3}$
for $1969 \sim 1975$ (Lyne et al. 1993).
For there are rare glitches occurred in this interval.
Later observations (Lyne et al. 2015) confirm previous results.

(b):The range of inclination angle is given by modeling the pulse
profile of the Crab pulsar (Dyks et al. 2003; Harding et al. 2008;
Watter et al. 2009; Du et al. 2012; Lyne et al. 2013).
\end{center}
\end{table}

\section{Spin-down of the Crab pulsar in the pulsar wind model}

\subsection{Description of the pulsar wind model}

In the pulsar wind model, the rotational energy is consumed by
magnetic dipole radiation and particle acceleration (Xu \& Qiao 2001).
The magnetic dipole radiation is related to the
perpendicular component of the magnetic moment ($\mu_{\perp}=\mu
\sin \alpha$), the power of magnetic dipole radiation is (Shapiro
\& Teukolsky 1983):
\begin{equation}
\dot{E_{\rm d}}=\frac{2 \mu^{2} \Omega^{4}}{3 c^{3}}
\sin^{2}\alpha ,
\label{Edotdipole}
\end{equation}
where $\Omega=2 \pi \nu$ is the angular speed of the pulsar. The
effect of parallel component of magnetic moment
($\mu_{\parallel}=\mu \cos \alpha$) is responsible for particles
acceleration (Ruderman \& Sutherland 1975). Acceleration gaps are
formed above the polar gap, primary particles are generated and
accelerated in these gaps. The energy taken away by these
particles dependents on the acceleration potential drop. The
corresponding rotational energy loss rate is (Xu \& Qiao 2001):
\begin{equation}
\dot{E_{\rm p}}=2 \pi r_{\rm p}^{2} c \rho_{\rm e} \Delta \phi ,
\label{Edotp}
\end{equation}
where $r_{\rm p}=R(R \Omega/c)^{1/2}$ is polar gap radius,
$\rho_{\rm e}=\kappa \rho_{\rm GJ}$ is the primary particle
density ($\rho_{\rm GJ}=\Omega B/(2 \pi c)$ is the
Goldreich-Julian charge density, Goldreich \& Julian 1969 ),
$\Delta \phi$ is the corresponding acceleration potential of the
acceleration region, and $\kappa$ is a coefficient related to
primary particle density which can be constrained by
observations\footnote{Primary particle density in the acceleration
gap is: $\rho_{\rm e}=|e|[n_{+}+n_{-}]$, but Goldreich-Julian
``charge'' density is: $\rho_{\rm GJ}=|e|[n_{+}-n_{-}]$ (Yue et
al. 2007). It is reasonable to infer that: $\kappa\geq1$.}. The
maximum acceleration potential for a rotating dipole is $\Delta
\Phi=\mu \Omega^{2}/c^{2}$ (Ruderman \& Sutherland 1975). Then
equation (\ref{Edotp}) can be rewritten as (considering that the
particle acceleration is mainly related to the parallel component
of magnetic moment):
\begin{equation}
\dot{E_{\rm p}}=\frac{2 \mu^{2} \Omega^{4}}{3c^{3}} 3 \kappa
\frac{\Delta \phi}{\Delta \Phi} \cos^{2} \alpha .
\label{EdotP}
\end{equation}
The magnetic dipole radiation and the outflow of particle wind may
contribute independently. Then the total rotational energy loss rate
is:
\begin{eqnarray}
\dot{E} &=& \dot{E_{\rm d}} + \dot{E_{\rm p}}
= \frac{2 \mu^{2}
\Omega^{4}}{3 c^3}(\sin^{2} \alpha + 3 \kappa \frac{\Delta \phi}{\Delta \Phi} \cos^{2} \alpha)
\nonumber\\
&=&\frac{2\mu^{2}\Omega^{4}}{3c^{3}}\eta,
\label{Edottotal}
\end{eqnarray}
where $\eta = \sin^2 \alpha + 3 \kappa \Delta \phi/{\Delta \Phi}
\cos^{2} \alpha$. The first and second items of expression $\eta$
are respectively for magnetic dipole and particle wind. If the
acceleration potential $\Delta \phi=0$, there are no particles
accelerated in the gap, the pulsar is just braked down by the
magnetic dipole radiation. The acceleration potential is model
dependent (Xu \& Qiao 2001; Wu et al. 2003). A particle density of
$\kappa$ times Goldreich-Julian charge density is considered.
Expressions of $\eta$ are given in Table
\ref{expressions} for various acceleration models\footnote{As shown in
Table \ref{expressions}, the models of VG(CR) and SCLF(I) are very
similar with each other. Crab parameters calculated in these
two models are also similar (shown later).}. The rotational
evolution of the pulsar can be written as:
\begin{equation}
-I \Omega \dot \Omega = \frac{2 \mu^{2} \Omega^{4}}{3 c^{3}}
\eta ,
\label{Edot}
\end{equation}
According equation (\ref{defn}) and (\ref{defm}), the braking index
and second braking index in the wind braking model are:
\begin{equation}
n = 3 + \frac{\Omega}{\eta} \frac{d \eta}{d \Omega}  ,
\label{nwind}
\end{equation}
\begin{equation}
m = 15 + 12 \frac{\Omega}{\eta} \frac{d \eta}{d \Omega} +
(\frac{\Omega}{\eta} \frac{d \eta}{d \Omega})^{2} +
\frac{\Omega^{2}}{\eta} \frac{d^{2} \eta}{d \Omega^{2}}.
\label{mwind}
\end{equation}
Equation (\ref{Edot}), (\ref{nwind}) and (\ref{mwind}) are all
functions of magnetic field B, inclination angle $\alpha$ and
particle density $\kappa$ ($\Omega$ and its derivatives are
observational inputs, see Table \ref{parameters}).

Taking these expressions of $\eta$ (Table \ref{expressions}) back
to equation (\ref{Edot}), it can be seen that: as the pulsar
spinning down ($\Omega$ decreases), the effect of particle wind
becomes stronger. If there are particles flowing out, it means
that there are still particle acceleration in magnetosphere (the
pulsar is still active). However, as pulsar spinning down, the
potential drop may not sustain the need to accelerate particles
(Ruderman \& Sutherland 1975). Besides, observations do not
support pulsars to evolve unceasingly in these pulsar wind models
(Young et al. 1999; Contopoulos \& Spitkovsky 2006). Therefore,
``pulsar death'' must be considered in modeling the long term
rotational evolution of the pulsar. As the spin angular speed
decreasing to the death value, the radio emission tends to stop.
And when pulsar is death, the pulsar is braked only by magnetic
dipole radiation. Here, the VG(CR) model is employed to show the
effect of pulsar death by introducing a piecewise function
deduction (details are shown in the appendix A).
\begin{scriptsize}
\begin{equation}\label{eqn_Edotdeath}
\eta^{\rm CR}_{\rm VG}
= \left\{
\begin{array}{l}
\sin^{2}\alpha + 4.96\times 10^{2} \kappa (1-\frac{\Omega_{\rm death}}{\Omega})
B_{12}^{-8/7}\Omega^{-15/7} \cos^{2}\alpha, \\
\mbox{\quad if $\Omega > \Omega_{\rm death}$} \\
\sin^{2}\alpha, \\
\mbox{\quad if $\Omega < \Omega_{\rm death}$.} \, \end{array} \right.
\end{equation}
\end{scriptsize}
The death period is defined as $\Omega_{\rm death}=2 \pi/ P_{\rm
death}$ (Contopoulos \& Spitkovsky 2006; Tong \& Xu 2012):
\begin{equation}\label{Pdeath}
P_{\rm death}=2.8 (\frac{B}{10^{12} \, \rm G})^{1/2} (\frac{V_{\rm
 gap}}{10^{12} \, \rm V})^{-1/2} \, \rm s .
 \end{equation}
$V_{\rm gap}$ can be viewed as the maximum acceleration potential
drop in the open field line regions and $V_{\rm gap} = 10^{13} \,
\rm V$ is chosen in the following calculations (Contopoulos \&
Spitkovsky 2006). It is introduced according to Contopoulos \&
Spitkovsky (2006) (see also later works by Li et al. 2012). In
this way, the effect of pulsar death can be incorporated in the
rotational energy loss rate. The effect of pulsar death is
discussed in more detail in section 2.5. For young pulsars
($\Omega \gg \Omega_{\rm death}$), the effect of pulsar death is
negligible. But for the long-term evolution of pulsars, especially
when the period approaches the death period, the effect of pulsar
death is significant.

\begin{table}
\scriptsize
 \begin{center}
 \caption{The Expressions of $\eta$ for nine particle acceleration models.}
 \label{expressions}
    \begin{tabular}{lll}
      \hline \hline
     No. & Acceleration model & $\eta$ \\
     \hline
     1 &  VG (CR) &  $\sin^2\alpha+ 4.96\times 10^2 \kappa B_{12}^{-8/7} \Omega^{-15/7} \cos^2\alpha$\\
     2 & VG (ICS) & $\sin^2\alpha+ 1.02\times 10^5 \kappa B_{12}^{-22/7} \Omega^{-13/7} \cos^2\alpha$\\
     3 & SCLF (II,CR) & $\sin^2\alpha+ 38 \kappa B_{12}^{-1} \Omega^{-7/4} \cos^2\alpha$\\
     4 & SCLF (II, ICS) & $\sin^2\alpha+ 2.3 \kappa B_{12}^{-22/13} \Omega^{-8/13} \cos^2\alpha$\\
     5 & SCLF (I)  & $\sin^2\alpha+ 9.8\times 10^2 \kappa B_{12}^{-8/7} \Omega^{-15/7} \cos^2\alpha$\\
     6 &  OG    & $\sin^2\alpha+ 2.25\times 10^5 \kappa B_{12}^{-12/7} \Omega^{-26/7} \cos^2\alpha$\\
     7 & CAP  & $\sin^2\alpha+ 54 \kappa B_{12}^{-1} \Omega^{-2} \cos^2\alpha$\\
     8 & NTVG (CR) & $\sin^{2} \alpha +13.7 \beta^{0.14} \kappa B_{12}^{-1} \Omega^{-1.76}
     \cos^{2}\alpha$\\
     9 & NTVG (ICS) & $\sin^{2} \alpha +69.6 \gamma^{-1} \kappa B_{12}^{-1} \Omega^{-1.88}
     \cos^{2}\alpha$ \\
      \hline
    \end{tabular}
  \flushleft
Notes: Particle density $\rho_{\rm e}=\kappa \rho_{\rm GJ}$ is
taken in all these models. $B_{\rm 12}$ is the magnetic field
strength in units of $10^{12} \,\rm G$.

(a): The models $1 \sim 5$ are based on the pulsar wind model of
Xu \& Qiao (2001). The VG and SCLF are respectively the
acceleration models: vacuum gap and space charge limit flow. In
the vacuum gap (Ruderman \& Sutherland 1975), curvature radiation
(CR) and inverse Compton scattering (ICS) are considered (Zhang et
al. 2000). In the SCLF case, regimes II and I are defined by field
saturated or not (Arons \& Scharlemann 1979; Harding \& Muslimov
1998). Three models are introduced: CR for SCLF model in regime
II, ICS for SCLF model in regime II, and the SCLF model in regime
I.

(b): The OG means the outer gap, is self-sustaining and limited by the
electron-positron pair produced by collisions between high-energy
photons (Zhang \& Cheng 1997). The modification by Wu et al. (2003)
is adopted.

(c): The CAP model is a phenomenological model of constant potential
$\Delta \phi=3 \times 10^{12} \rm V$ (Yue et al. 2007).

(d): The NTVG shorts for near threshold vacuum gap. In
this model, the electron-positron pair produced at or near the
kinematic threshold $\hbar \omega=2mc^{2}/\sin \theta$ because of
superstrong surface magnetic field (Gil \& Melikidze 2002).
$\beta=52$  is taken in the CR case, and $\gamma=14$ is taken
in the ICS case (Abrahams \& Shapiro 1991; Wu et al. 2003).
\end{center}
   \end{table}

\subsection{Understanding the braking index of the Crab pulsar}

Parameters of the Crab pulsar are calculated in this section. The
VG(CR) model is taken as an example to show the calculation
process. For pulsars with the observed $\nu$, $\dot{\nu}$,
$\ddot{\nu}$ and $\stackrel{...}{\nu}$, their observational
braking index $n$ and second braking index $m$ can be get by
equation (\ref{defn}) and (\ref{defm}). Pulsar parameters such as
magnetic field, inclination angle, and particle density can be
calculated by these three equations (\ref{Edot}), (\ref{nwind})
and (\ref{mwind}). The primary particle density may be $10^{3}\sim
10^{4}$ times of Goldreich-Julian charge density (Yue et al 2007).
According the observational range of inclination angle and the
characteristic magnetic field (which can be viewed as order of
magnitude estimation of the true magnetic field), the range of
particle density $\kappa$ can be limited by equation (\ref{Edot})
and (\ref{nwind}). In our calculation, $\kappa=10^3$ is
adopted\footnote{We should note that the $\kappa$ is related to
the primary particles in the acceleration gap but not the total
out-flow particles.}, the inclination angle $\alpha$ and magnetic
field $B$ are about $55^{\circ} $ and $8.1 \times 10^{12} \, \rm
G$ which are consistent with observational constraints. The second
braking index calculated by equation (\ref{mwind}) is 10.96, which
is roughly consistent with the observed value 10.15 (Lyne et al.
1993), considering the observational uncertaintities (Lyne et al.
1993, 2015).

\begin{figure}
\centering
\includegraphics[width=0.45\textwidth]{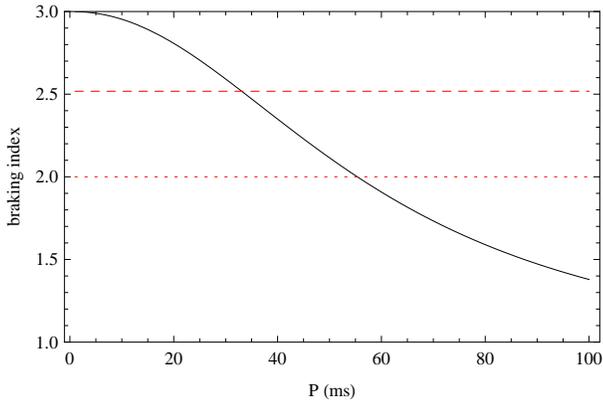}
\caption{The braking index of the Crab pulsar as function of spin
period in the VG(CR) model. The dashed line is observational
braking index of $2.51$. The dotted line is braking index in
the transition point ($n=2$).}
\label{fignp}
\end{figure}

The magnetic field and inclination angle are assumed to be
constant in the pulsar wind model. The particle density decreases
in long-term evolution of pulsars but remains unchanged in short
duration at early time. Figure \ref{fignp} shows the braking index
of the Crab pulsar as a function of spin period. As the pulsar
period increases, the braking index decreases from 3 to 1. It
illustrates that the Crab pulsar is initially braked by magnetic
dipole ($n \rightarrow 3$) and then by the particle wind ($n
\rightarrow 1$). We take $ n = 2$ as the transition point (see
Figure \ref{figevo&short}, at $n=2$, the slop of the curve is $0$)
and the corresponding spin period is about $55 \, \rm ms$. Since
Table \ref{parameters} has shown the timing observations of the
Crab pulsar in 1969 (its age is 915 years old by then), its
initial period is $P_0 \approx 18.8 \, \rm ms$ in AD 1054 by
integrating equation (\ref{Edot}). The calculations of all
acceleration models are shown in Table \ref{tab_calculations}. As
shown in this table, the particle densities are different in
different models, but are all more than $100$ times of $\rho_{\rm
GJ}$. The inclination angles is within the observational range.
The initial periods are all about $19 \, \rm ms$.

\begin{table}
\scriptsize
\begin{center}
\caption{Parameters of the Crab pulsar calculated in various acceleration
models.} \label{tab_calculations}
\begin{tabular}{lllllllll}
\hline \hline
 Models & $\kappa$ & $\alpha$ & $B_{\rm 12}$ & $P_{\rm 0}$ &
$P_{\rm t}$ & $P_{\rm d}$ & $T_{\rm t}$ & $T_{\rm d}$ \\
 & $10^{3}$ & $^{\circ}$ &  & $\rm ms$ & $\rm ms$ & $\rm s$ & $10^3 \,\rm yr$ & $10^5 \,\rm yr$ \\
\hline VG(CR) & $1$ & $55$ & $8.1$ & $18.8$ & $55$ & $2.52$ &
$2.77$ & $0.78$ \\
VG(ICS) & $0.1$ & $60$ & $7.5$ & $18.9$ & $63$ & $2.58$ & $3.50$
& $1.67$ \\
SCLF(II CR) & $2$ & $58$ & $7.5$ & $18.9$ & $68$ & $2.59$ &
$3.93$ & $2.24$ \\
SCLF(II ICS) & $0.6$ & $44$ & $5.0$ & $19.2$ & $-$ & $3.16$ & $-$
& $30.1$ \\
SCLF(I) & $0.6$ & $57$ & $7.8$ & $18.8$ & $55$ & $2.45$ & $2.77$ &
$0.78$ \\
OG & $20$ & $59$ & $8.2$ & $18.5$ & $42$ & $2.47$ & $1.63$ &
$0.07$ \\
CAP & $3$ & $52$ & $8.3$ & $18.9$ & $58$ & $2.46$ & $3.08$ &
$1.14$ \\
NTVG(CR) & $3$ & $57$ & $7.7$ & $19.5$ & $67$ & $2.56$ & $4.10$ &
$2.52$ \\
NTVG(ICS) & $30$ & $54$ & $7.5$ & $19.4$ & $62$ & $2.58$ & $3.58$
& $1.81$ \\
\hline
\end{tabular}
\flushleft

Notes:

(a): $\kappa$ is the coefficient of primary particle density
$\rho_{\rm e}=\kappa \rho_{\rm GJ}$.

(b):$\alpha$ and $B_{\rm 12}$ are respectively the inclination angle
and the magnetic field.

(c): $P_{\rm 0}$ is the initial period.

(d):$P_{\rm t}$  and $T_{\rm t}$ are the transition period and corresponding age.
At this transition point the braking index is $2$. Because the minimum braking index of model
SCLF(ICS) is $2.4$, there is no such a transition point, which can
be seen in the evolution in $P-\dot{P}$ diagram (see Figure \ref{figevo&models}).

(e): $P_{\rm d}$ and $T_{\rm d}$ are the period and age
when pulsar stops radio emission. The death period is calculated
by equation (\ref{Pdeath}).

\end{center}
\end{table}

\subsection{An increasing particle density results in a lower braking index during glitches}

A braking index $2.3$ has been monitored after the removal of the
effects of glitches during an epoch when there are many glitches
occured (Lyne et al. 2015). It is different from the long term
underlying braking index $2.51$ of the Crab pulsar. Previously,
Wang et al. (2012) have measured two different values of braking
index $n=2.45$ and $ n=2.57$ between glitches. They attribute it
to the effect of varying particle wind strength (Wang et al.
2012). It is generally accepted that glitch is caused by the inner
effect but may lead to some effect in the outer
magnetosphere\footnote{Glitch induced magnetospheric activities
are very common in the case of magnetars (Dib \& Kaspi 2014).},
for example: the changing of primary particle density during this
epoch. In the pulsar wind model, it can be denoted by the changing
coefficient of particle density $\kappa=\kappa(t)$. Then equation
(\ref{nwind}) should be modified:
\begin{equation}\label{nkappa}
n=3 + \frac{\Omega}{\eta} \frac{d \eta}{d \Omega} +
\frac{\kappa}{\eta} \frac{d \eta}{d \kappa} \frac{\Omega}{\dot
\Omega} \frac{\dot \kappa}{\kappa}.
\end{equation}
Equation (\ref{nkappa}) can be rewritten as:
\begin{equation}
n=3 + \frac{\Omega}{\eta} \frac{d \eta}{d \Omega}-
\frac{\kappa}{\eta} \frac{d \eta}{d \kappa}
\frac{\tau_{c}}{\tau_{\kappa}},
\label{ntauk}
\end{equation}
where $\tau_{c}=- \frac{\Omega}{2 \dot{\Omega}}$ is the
characteristic age, $\tau_{\kappa}=\frac{\kappa}{2 \dot \kappa}$
can be viewed as the typical time scale of the changing particle
density. When the magnetosphere is in equilibrium (i.e., when the
glitch activities of the pulsar is not very active), the time
scale of particle density variation ($\tau_{\kappa}$) may be very
large (i.e., larger than the characteristic age $\tau_{c}$). The
third term in equation (\ref{ntauk}) does not contribute. But, if
they are comparable with each other, the braking index changes
substantially. In other words, when the outflow particle density
increases and the effect of particle wind becomes stronger, means
$\tau_{\kappa}>0$ or $\dot{\kappa}>0$, the braking index will be
smaller than $2.51$. But when the outflow particle density
decreases ($\tau_{\kappa}<0$ or $\dot{\kappa}<0$), the braking
index will be larger than $2.51$. As the out-flow particle density
tending to a certain value (larger or smaller than previous
value), the braking index tends to its underlying value ($2.51$).
Generally, the increased (or decreased) component of the outflow
particle density (may be $0.1\%$) is so much small that it can be
ignored.

\begin{figure}
\centering
\includegraphics[width=0.45\textwidth]{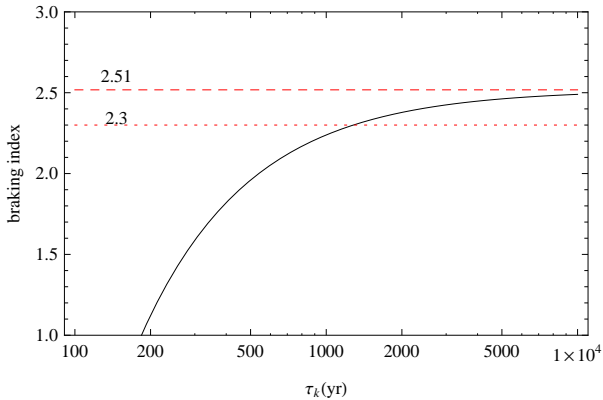}
\caption{The braking index of the Crab pulsar as function of
$\tau_{\kappa}$ in the VG(CR) model. The dashed line is the
underlying braking index $2.51$. The dotted line is the smaller
braking index $2.3$ measured during glitch activities (Lyne et al.
2015, Figure 7 there).} \label{fignk}
\end{figure}

Figure \ref{fignk} shows the braking index as function of
$\tau_{\kappa}$ for the Crab pulsar in the VG(CR) model. In order
to better understand the observations of $n<2.51$ (Lyne et al.
2015), we consider the $\tau_{\kappa}>0$ only. As we can see in
this figure, the braking index is insensitive to $\tau_{\kappa}$
when it is larger than $10^{4}\,\rm yr$. But, when it is
comparable with the characteristic age (about $10^{3}$ yr), the
braking index decreases sharply. When braking index is $2.3$, the
changing rate of the particle density is $\dot{\kappa}\approx
1.2\times 10^{-8} \, \rm s^{-1}$. In this interval (from MJD 51000
to MJD 53000, about $5$ years), the particle density has changed
by $2 \, \rho_{\rm GJ}$. The change of particle density will
result in changes of radiation energy loss rate and period
derivative. Their changes respectively are: $\delta \dot{E}\approx
2\times 10^{35} \,\rm erg/s$ and $\delta\dot{P}\approx1.88\times
10^{-16} \, \rm s/s$.

\begin{figure}
\centering
\includegraphics[width=0.45 \textwidth]{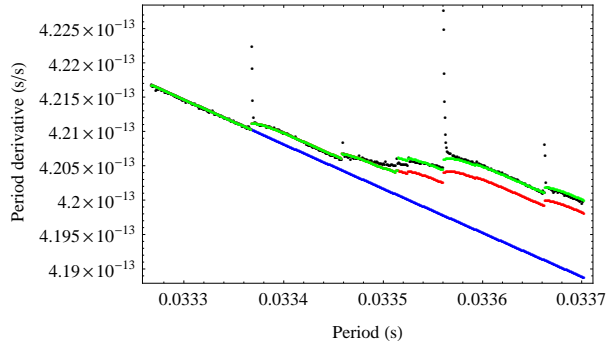}
\caption{Evolution of the Crab pulsar in $P-\dot{P}$ diagram in VG(CR) model
compared with observations. The black points are timing results
from Jodrell Bank monthly ephemeris. The blue points are the rotational
evolution of Crab pulsar in the VG(CR) model. The red ones are obtained
by adding the effect of glitches (Lyne et al. 2015, Table 3 there). And
the green ones are the summation of red points and a factor
$\delta\dot{P}\approx1.88\times 10^{-16} \, \rm s/s$ (from MJD 51804.75 afterwards),
which may be caused by the change of out-flow particle density.
At first, the black, blue, red, and green points are coincide with each other.
(Notes: The fitting does not model the transient process caused by
glitches.)}
 \label{figob&model}
\end{figure}

The pulsar wind model can be tested with observations (Jodrell
Bank monthly ephemeris of the Crab
pulsar\footnote{http://www.jb.man.ac.uk/pulsar/crab.html, data
from 1982 February 15 (MJD 45015) to 2014 October 15 (MJD 56945)
are used here.}). The effect of glitches is not taken into
consideration in the pulsar wind model. The changes of period and
period derivative caused by glitches should be added (Lyne et al.
2015). Figure \ref{figob&model} is the model calculation compared
with observations (without modelling the transient variation
caused by glitches). When the amplitude of glitches is relatively
small, the change of particle density is not obvious, then the
model calculations (the blue points) fit well with observations
(the black points). However, when the amplitude of glitches is
large, the effect of glitches must be added (the red points).
Moreover, the effect of particle density change should be taken
into consideration. A factor $\delta\dot{P}=1.88\times10^{-16}\,
\rm s/s$ (which is calculated above) is added to the red points
(from MJD 51804.75 afterwards), shown as the green points. The
green points can match the general trend of the observations.

\subsection{Long term evolution of the Crab pulsar}

\begin{figure}
\begin{minipage}{0.45\textwidth}
 \includegraphics[width=0.95\textwidth]{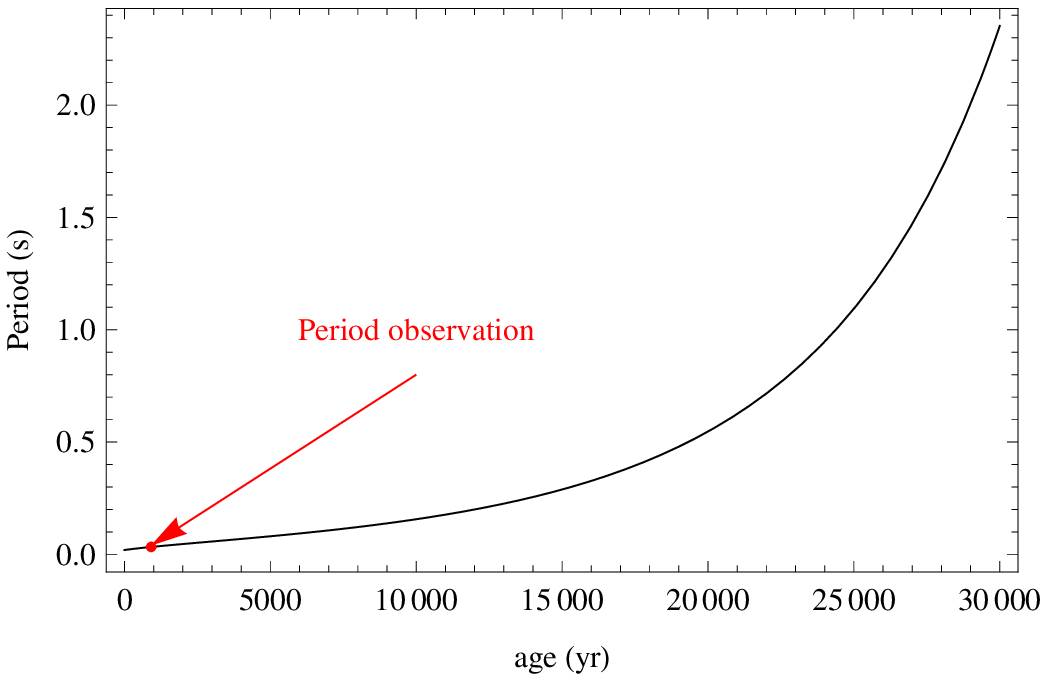}
\end{minipage}
\begin{minipage}{0.45\textwidth}
 \includegraphics[width=0.95\textwidth]{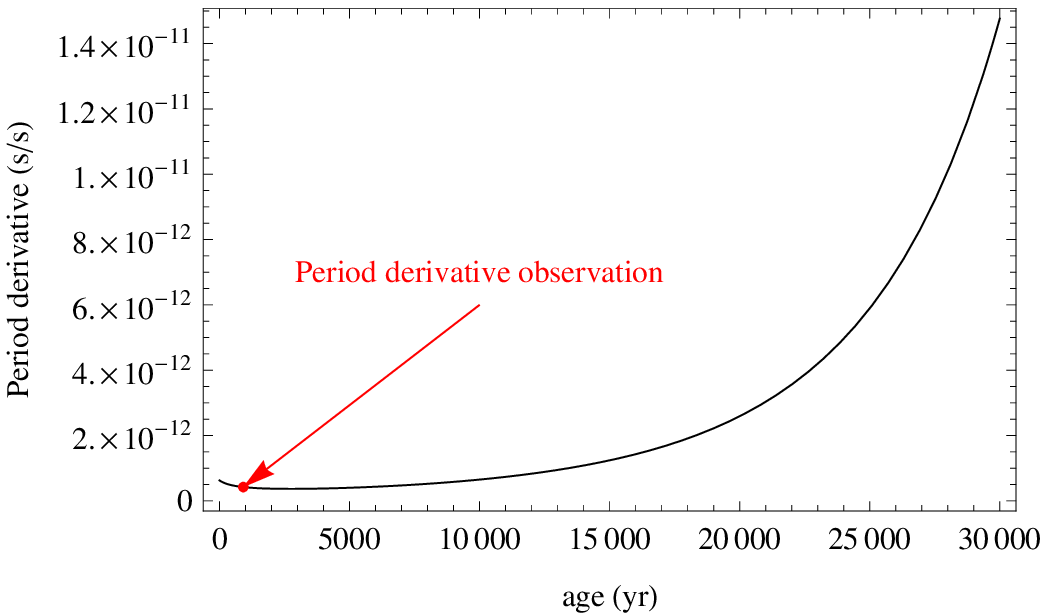}
\end{minipage}\\
\begin{minipage}{0.45\textwidth}
 \includegraphics[width=0.95\textwidth]{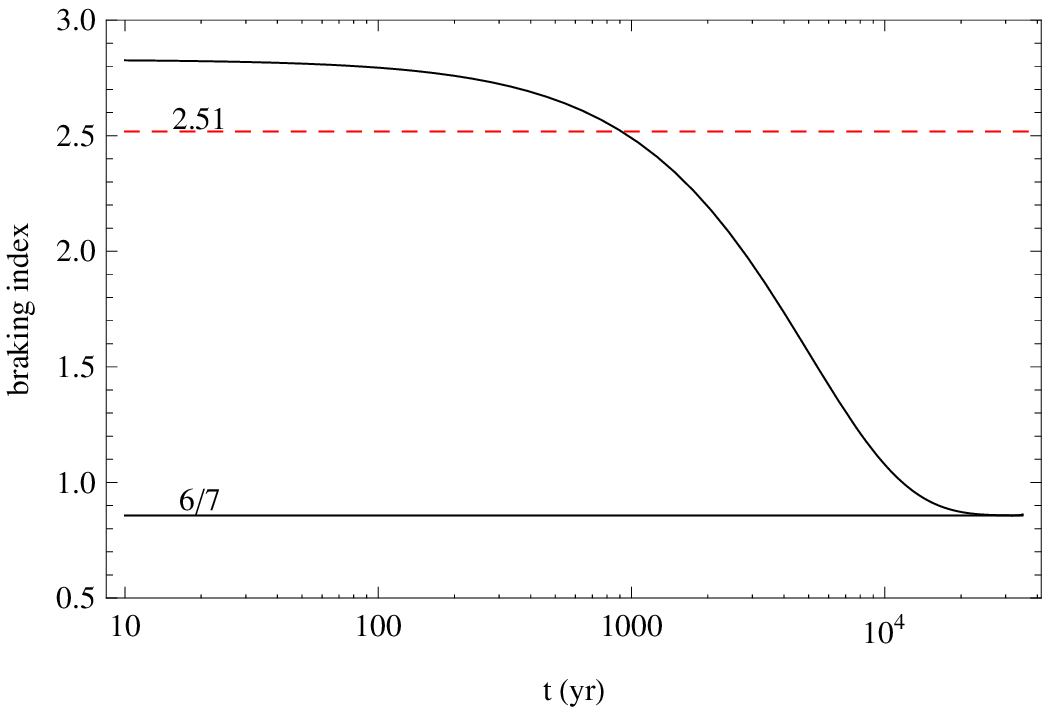}
\end{minipage}
\begin{minipage}{0.45\textwidth}
 \includegraphics[width=0.95\textwidth]{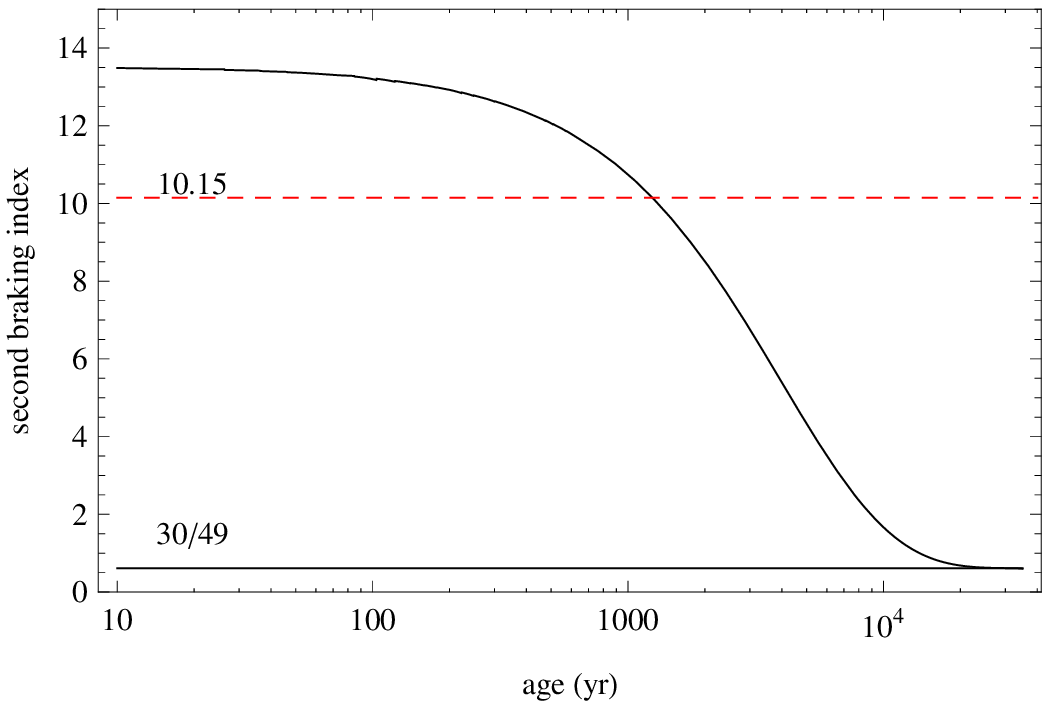}
\end{minipage}
\caption[]{Rotational evolution of the Crab pulsar from birth to 30 thousands
years in the VG(CR) model. The first and second panels are
respectively the period and period derivative
evolution with time. The points are observations of period and its
derivative (Lyne et al. 1993). The third and fourth panels
are respectively the braking index and second braking index
evolution with time. The dashed lines are observed values and
the solid lines are respectively the minimum braking index and
second braking index in the VG(CR) model.}
\label{figparameters}
\end{figure}

\begin{figure}
\centering
\includegraphics[width=0.45\textwidth]{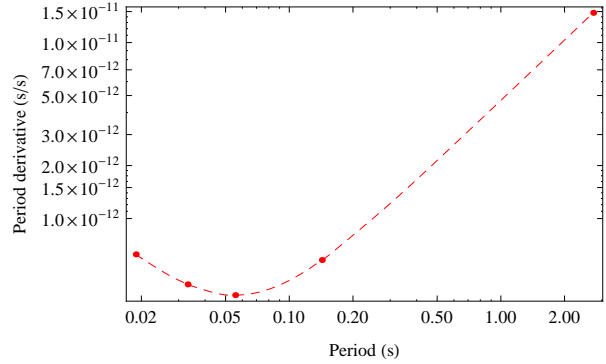}
\caption{Evolution of the Crab pulsar in the $P$-$\dot{P}$
diagram. Five points are chosen when the pulsar is: $1$, $915$,
$2771$, $9000$ and $30000$ years old.} \label{figevo&short}
\end{figure}

The pulsar period and its fist derivative evolution with time in
the VG(CR) model are shown in the first and second panel of Figure
\ref{figparameters}. The curves evolve slowly in its early age and
then rise sharply, indicating that the Crab pulsar is mainly
braked by magnetic dipole radiation firstly and then mainly by
particle wind. The transition period is $55 \,\rm ms$ and age of
the Crab pulsar will be $2771$ calculated in the VG(CR) model (see
Table \ref{tab_calculations}). The braking index evolution with
time in the VG(CR) model is shown in the third panel of Figure
\ref{figparameters}. Braking index decreases from $3$ to $6/7$
--different minimum braking indices for different acceleration
models. The bottom panel of Figure \ref{figparameters} shows the
second braking index evolution with time. The curve gradually
decreases from $15$ to 30/49. Figure \ref{figevo&short} shows the
evolution of the Crab pulsar in $P-\dot{P}$ diagram from birth to
its $30000$ years old. These points are respectively taken when
the Crab pulsar is: $1$, $915$, $2771$, $9000$ and $30000$ years
old. Period derivative evolves with period by a function of
$\dot{P} \propto P^{2-n}$ (Espinoza et al. 2011). In the log-log
plot, the evolution curve evolves with a slop ($2-n$). As the
braking index falling from $3$ to $1$, the pulsar evolve from
magnetic dipole radiation dominated case (the left points) to the
particle wind dominated case (the right points) and the bottom is
the transition point ($n=2$). Clearly, the curve evolves more
sharply after the transition point. The asymptotic behaviors are
discussed in the appendix B.

\subsection{The effect of pulsar death}

\begin{figure}
\centering
\includegraphics[width=0.45\textwidth]{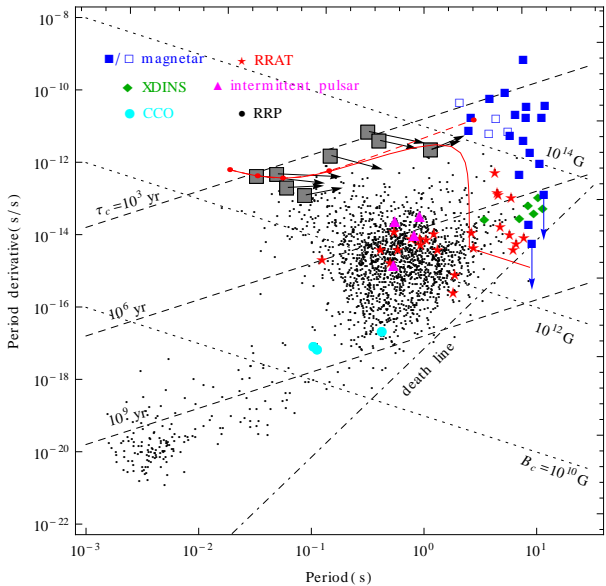}
\caption{Rotational evolution of the Crab pulsar in the VG(CR)
model. The dashed curve is the evolution  without considering
pulsar death. The solid line is the one considering pulsar death.
The $P-\dot{P}$ diagram has listed all known radio pulsars,
magnetars, and millisecond pulsars (updated from Figure 1 in Tong
\& Wang 2014). The dot-dashed line represents the fiducial death
line (Ruderman \& Sutherland 1975). Pulsars with meaningful
braking indices (Espinoza et al. 2011; Lyne et al. 2015) are
defined by large square. The arrows represent their evolution
directions and each arrow indicate its motion in the next 10000
\,yr (For the Vela pulsar, 20000 yr is used; 5000 yr for PSR
J1846-0258 and PSR J1119-6127).} \label{figevo&long}
\end{figure}

As shown is the Figure \ref{figevo&short}, the line evolves up
right with a slope about $1$ after the transition point just like
PSR J1734-3333 whose braking index is $0.9\pm0.2$ (Espinoza et al.
2011). If the pulsar continually evolve in this trend, it would
not dead and might end itself as a magnetar. It will be hard to
explain the observations for: (1) The number of magnetars are
smaller than radio pulsars and most of the pulsars occupy the
central region in the $P$-$\dot{P}$ diagram (see Figure
\ref{figevo&long}); (2) Only a small portion of magnetars are
radio emitters (Olausen \& Kaspi 2014); (3) The properties of the
Crab pulsar, PSR J1734-3333 and other high magnetic field pulsars
are significantly different from magnetars (Ng \& Kaspi 2011). As
pulsars slow-down, the density of outflow particles may decrease
and the effect of particle wind will recede (for the SCLF cases,
the number of particles (or $\kappa$) may close to constant, but
the particle density still decreases because a reducing
Goldreich-Julian charge density). Therefore, the effect of pulsar
death must be included in modeling the long-term rotational
evolution of pulsars.

The essential condition of radio emission for a pulsar is its pair
production (Sturock 1971; Ruderman \& Sutherland 1975). The so
called pulsar ``death'' means the stopping of pulsar emission
(Ruderman \& Sutherland 1975; Chen \& Ruderman 1993; Zhang et al.
2000; Contopoulos \& Spitkovsky 2006). The death defined by
Ruderman \& Sutherland (1975) is that: a pulsar dead when the
maximum potential drop $\Delta \Phi$ available from the pulsar is
smaller than the acceleration potential needed to accelerate
particles. Because this death line is model dependent, Chen \&
Ruderman (1993) proposed death ``valley'' by modifying the
boundary conditions. And this death ``valley'' was updated by
considering different particle acceleration and photon radiation
processes (Zhang et al. 2000). However these works are just to
separate pulsars with radio emission from those without.
Contopoulos \& Spitkovsky (2006) included the effect of pulsar
death when modeling the rotational evolution of pulsars. However,
the braking indices there are always larger than three.
Considering particle acceleration and pulsar death simultaneously
may give a comprehensive interpretation for both short and
long-term rotational evolution of pulsars.

\begin{figure}
\centering
\includegraphics[width=0.45\textwidth]{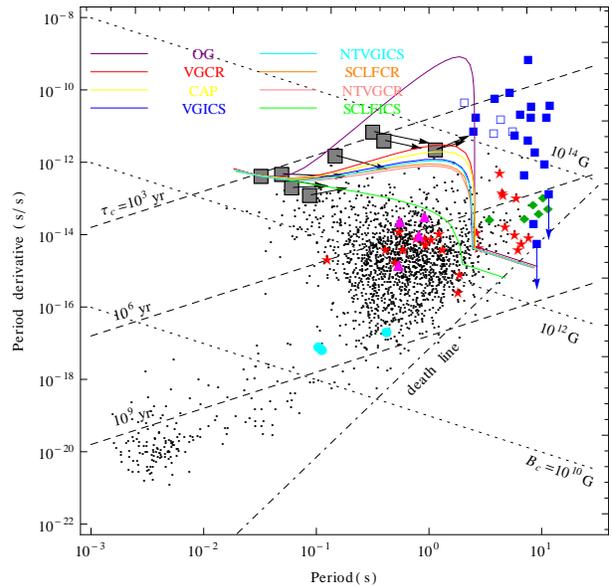}
\caption{Rotational evolution of the Crab pulsar in all acceleration models.
Because the VG(CR) model and SCLF(I) model are very similar, their
lines are coincident. Only the line in the VG(CR) model is plotted.
See also Figure \ref{figevo&long} for explanations.}
\label{figevo&models}
\end{figure}

The visual explanation for pulsar death in the pulsar wind model
is that the particles in the magnetosphere are exhausted and there
are no primary particles generated. For the physical
understanding, we can refer to the equation (\ref{Edottotal}),
expressions $\Delta{\Phi}\propto \Omega^{2}$ and $\Delta{\phi}$
(i.e., $\Delta{\phi}_{\rm VGCR}=2.8\times10^{13}R_{\rm
6}^{2/7}B_{12}^{-1/7}\Omega^{-1/7} \rm \,V$, Xu \& Qiao 2001),
with the pulsar spinning down, the maximum potential of the pulsar
($\Delta{\Phi}$) drops, when the maximum potential can not meet
the need to accelerate particles ($\Delta{\phi}>\Delta{\Phi}$),
the pulsar death. Besides, the acceleration potential is weakly
dependent on the $\Omega$, that is why we can make calculations
under the constant acceleration potential (the CAP model) (Yue et
al. 2007). Figure \ref{figevo&long} shows the long-term evolutions
of the Crab pulsar in the VG(CR) model. The dashed curve is the
line shown in Figure \ref{figevo&short} and the solid is the one
with pulsar death (according to equation (\ref{eqn_Edotdeath})).
These two lines evolve similar at the early age, and divide
obviously after the transition point. The second line falls down
afterwards and moves down-right with a slop about $-1$ finally.
The initial period and transition period can be calculated
according the second line, which are same with the values given by
the first line within precision. Indicating that the effect of
pulsar death can be ignored in the early time. As mentioned
before, the particle density decreases as pulsar spin-down. The
gap between these two lines is caused by the reducing of particle
density. When the spin period of Crab approaches to the death
period $2.52 \, \rm s$, the radio emission tends to stop and the
slop of the second line is very large. Crab evolution in various
acceleration models are given in Figure \ref{figevo&models}. The
bottom line is the evolution curve in SCLF(ICS) model. There is no
transition point, indicating that the particle wind effect of this
model is very weak. The top line is the evolution of the Crab
pulsar in the OG model. This model is for high-energy radiation,
the death of this model may differ from radio emission models. The
results here for the OG model are just crude approximations to
show the effect of pulsar death. Evolution of the Crab pulsar in
other models are similar with each other.

\section{Discussions}

In the magnetic dipole radiation model, a pulsar is assumed as a
magnetic dipole rotating in vacuum. It is just a low-order
approximation of the true magnetosphere, and there must be
particles acceleration and generation  process so that the radio
emission can be received. The actual magnetosphere surrounding the
pulsar must be filled with plasma and corotate with pulsar because
of magnetic freeze. According the special relativity, the
corotation magnetosphere can not infinitely extend but only exist
within the light cylinder. The magnetosphere is divided into two
parts by the light cylinder: the closed magnetic field lines and
the open magnetic field lines (including polar gap and outer gap).
In the open field line region, particles are accelerated, generate
radiation and flow out (thus taken away some of the rotational
energy of the central neutron star). In the pulsar wind model, the
pulsar is braked down by magnetic dipole radiation and particle
wind. But in the electric current loss model, the electric current
flow generates a spin-down torque (Harding et al. 1999; Beskin et
al. 2007). Based on the same magnetospheric physics, the pulsar
wind model and the current loss case are just different point of
views, will produce the same results (e.g. Section 3.2 in Tong et
al. 2013). And they can be used to check each other. It is
meaningful to take advantage of the existing information and
explain more observations. The particle wind worked on pulsar
spin-down was first proven by observations of intermittent pulsar
PSR B1931+24 whose radio emission switches between ``on'' and
``of'' state (Kramer et al. 2006). Corresponding calculations for
intermittent pulsars in pulsar wind model were made by Li et al.
(2014).

As one of the best observed pulsar, observations of the Crab
pulsar are relatively comprehensive (see Table \ref{parameters}
$\&$ timing results from Jodrell Bank website). It will be the
best target to test radiation models. The pulsar parameters can be
calculated by equations (\ref{Edot}), (\ref{nwind}) and
(\ref{mwind}) when model dependent acceleration potential (or
$\eta$) is given (see Table \ref{expressions}). The
Goldreich-Julian charge density is taken as the primary particle
density in these models of Xu \& Qiao (2001) and Wu et al (2003).
However, Yue et al. (2007) constrained the primary particle
density by observations of several pulsars and predicted that it
should be $10^{3} \sim 10^{4}$ times of Goldreich-Julian charge
density. In our calculations, the primary particle density is at
least $100$ times of Goldreich-Julian charge density. In the
VG(CR) model, a particle density $10^{3} \, \vert \rho_{\rm GJ}
\vert$ is chosen. When applying the pulsar wind model to
intermittent pulsars, a Goldreich-Julian charge density is assumed
(Li et al. 2014). As the pulsar spin-down, its out-flow particle
density decreases. In the $P-\dot{P}$ diagram, these intermittent
pulsars are on the right of those young pulsars (see Figure
\ref{figevo&long}). This may explain the different particle
density in the Crab pulsar and intermittent pulsars.

Other studies also found the need for a high particle density in
the pulsar magnetosphere, e.g., when modeling the eclipse of the
double pulsar system, giant pulses of the Crab pulsar, and
magnetar X-ray spectra (Lyutikov 2008 and references therein),
optical/ultraviolet excess of X-ray dim isolated neutron stars
(Tong et al. 2010, 2011). Magnetohydrodynamical simulations of the
Crab nebula also require a much higher electron flow from the
magnetosphere (Buhler \& Blandford 2014 and references therein).
Besides, a super Goldreich-Julian current has been proposed in a
series theoretical works about the pairs creations in the
acceleration gap (Timokhin \& Arons, 2013; Zanotti et al. 2012;
Hirotani 2006). Combined with the results in this paper, it is
possible that there are high density plasmas (e.g., $10^3$-$10^4$
times the Goldreich-Julian density) in both open and closed field
line regions in the pulsar magnetosphere.

\subsection{Applications to other sources}

Here the updated pulsar wind model is employed to compare with the
observations of the Crab pulsar. It can also be applied to other
sources. For eight pulsars with meaningful braking index measured
(Espinoza et al. 2011; Lyne et al. 2015), their magnetospheric
parameters can also be calculated, including magnetic field,
inclination angle, particle density etc. But the higher second
frequency derivatives of some pulsars (Hobbs et al. 2010) may be
attributed to the fluctuation of magnetosphere (Contopoulos 2007).
During this process, more information will be helpful to constrain
the model parameters, e.g., inclination angle, second braking
index. Their evolution on the $P-\dot{P}$ diagram can be
calculated, which will be similar to Figure \ref{figevo&long}. The
pulsar PSR J1734$-$3333 has braking index $n=0.9\pm 0.2$ (Espinoza
et al. 2011). Its rotational energy loss rate will be dominated by
the particle wind. Also from Figure \ref{figevo&long}, the braking
indices of young pulsars will naturally be divided into two
groups. Braking indices of pulsars in the first group are close to
three, their evolution directions are down to the right in the
$P-\dot{P}$ diagram ($\dot{P}\propto P^{2-n}\sim P^{-1}$). Braking
indices in the second group are close to one and their evolution
directions are up to the right ($\dot{P}\propto P^{2-n}\sim
P^{1}$). Generally speaking, sources in the second group will be
older than the ones in the first group. At present, there are some
hints for the evolution of pulsar braking index (Espinoza 2013).
Future more observations will deepen our understanding on this
aspect.

Glitch is a phenomenon of the pulsar suddenly spinning up and then
slowing down by a larger rate. In the pulsar wind model, the lower
braking index $2.3$ during glitches (of the Crab pulsar) is due to
the larger out-flow particle density. The increasing particle
density is a possible performance associated with glitches.
Glitches induced magnetospheric activities are very common in
magnetars (Kaspi et al. 2003; Dib \& Kaspi 2014). For the high
magnetic field pulsar PSR J1846$-$0258, it also shows a smaller
braking index after glitch (Livingstone et al. 2011). The cause of
a smaller braking index may be similar to the Crab pulsar case.
Due to a larger particle outflow, the pulsar should experience
some net spindown after glitch. This has already been observed
(Livingstone et al. 2010). Some primary calculations for PSR
J1846$-$0258 has been done previously (Tong 2014, based on a wind
braking model designed for magnetars). The smaller braking index
of PSR J1846$-$0258 after glitch is consistent with the results
here.

From Figure \ref{figevo&long} and Figure \ref{figevo&models}, the pulsar will go up-right in the $P-\dot{P}$
diagram in the wind braking dominated case. Therefore, for pulsars in the up right corner of the $P-\dot{P}$
diagram (e.g., high magnetic field pulsars and magnetars), their true dipole magnetic field may be much lower
than the characteristic magnetic field. For magnetar SGR 1806$-$20, its characteristic magnetic field
is about $5\times 10^{15} \,\rm G$ (Tong et al. 2013). This dipole magnetic field
is too high to understand comfortably (Vigano et al. 2013). Using the physical braking mechanism in this paper,
the dipole magnetic field of SGR 1806$-$20 will be much lower.
In the late stage of a pulsar, it will bent and go downward in the $P-\dot{P}$ diagram. The Crab pulsar and other
pulsars will not evolve into the cluster of magnetars. Furthermore, for magnetars, they also will go downward
in the $P-\dot{P}$ diagram when they are aged enough. This may corresponds to the case of low magnetic field
magnetars (Tong \& Xu 2012).

\section{Conclusions}

An updated pulsar wind model which considered the effect of
particle density and pulsar death is developed. It is employed to
calculate parameters and simulate the evolution of the Crab
pulsar. The braking index $n<3$ is the combined effect of magnetic
dipole radiation and particle wind. The lower braking indices
measured between glitches are caused by the increasing particle
density. By adding the glitch parameters and a change of period
derivative caused by the change of particle density, the
theoretical evolution curve is well fitted with observations
(Figure \ref{figob&model}). This may be viewed as glitch induced
magnetospheric activities in normal pulsars. Giving the model
dependent acceleration potential drop, the magnetic field,
inclination angle and particle density of the Crab pulsar can be
calculated. The evolution of braking index and the Crab pulsar in
$P-\dot{P}$ diagram can be obtained. In the $P-\dot{P}$ diagram,
the Crab pulsar will evolve towards the death valley, not to the
cluster of magnetars (Figure \ref{figevo&long}). Different
acceleration models are also considered. The possible application
of the present model to other sources (pulsars with braking index
measured, and the magnetar population) is also mentioned.

\section*{Acknowledgments}

The authors would like to thank the Referee Barsukov D.P. for
helpful comments and R.X.Xu, J.B.Wang for discussions. H.Tong is
supported Xinjiang Bairen project, West Light Foundation of CAS
(LHXZ201201), Qing Cu Hui of CAS, and 973 Program (2015CB857100).

\appendix

\section{Consideration of pulsar ``death''}

The treatment of Contopoulos \& Spitkovsky (2006) is employed in
the following. Assuming that the open field line regions rotate at
a angular velocity $\Omega_{\rm open}$ but not the angular
velocity of the neutron star\footnote{In fact, particles in the
acceleration gap rotate with a angular speed between the spinning
speed of pulsar surface and the open field lines (Ruderman \&
Sutherland 1975). We use the angular speed of the open field lines
as the boundary condition to describe the pulsar death because the
acceleration potential is insensitive to it in this pulsar wind
model.}.
\begin{equation}
\Omega_{\rm open}=\Omega-\Omega_{\rm death}
\end{equation}
The particle density in this acceleration gap is $\kappa$ times
the Goldreich-Julian charge density:
\begin{equation}
\rho_{\rm e} \approx \kappa \frac{\Omega_{\rm {open}} B}{2 \pi c},
\end{equation}
The energy taken away by the out-flowing  particles is:
\begin{equation}
\dot{E_{\rm p}}=2 \pi r^{2}_{\rm p} c \rho_{\rm e} \Delta \phi.
\end{equation}
The perpendicular component and parallel component of the magnetic
moment are respectively related to the magnetic dipole radiation
and particle acceleration. The rotation energy loss rate can be
written as:
\begin{eqnarray}
\dot{E}&=&\dot{E_{\rm d}}+ \dot{E_{\rm p}}
= \frac{2 \mu^{2} \Omega^{4}}{3 c^3}(\sin^{2} \alpha + 3 \kappa (1-\frac{\Omega_{\rm
death}}{\Omega}) \frac{\Delta \phi}{\Delta \Phi} \cos^{2} \alpha)
\nonumber \\
&=&\frac{2 \mu^{2} \Omega^{4}}{3 c^3} \eta,
\end{eqnarray}
with the expression of $\eta$
\begin{equation}
\eta = \sin^2 \alpha + 3 \kappa(1-\frac{\Omega_{\rm
death}}{\Omega}) \frac{\Delta \phi}{\Delta \Phi} \cos^{2} \alpha.
\end{equation}
For a constant acceleration potential $\Delta \phi=3 \times 10^{12}
\rm V$, correspondingly,
\begin{equation}
\eta = \sin^2 \alpha + 54 \kappa (1-\frac{\Omega_{\rm
death}}{\Omega})R^{3}_{\rm 6} B^{-1}_{\rm 12} \Omega^{-2}
\cos^{2} \alpha.
\end{equation}
For the physical acceleration models, $\eta$ is model dependent. Its
expression in the VG(CR) case  is calculated in the following.
The potential drop of open field lines is (Ruderman \& Sutherland 1975):
\begin{equation}
\Delta \phi= \frac{\Omega_{\rm {open}} B}{c} h^{2},
\end{equation}
where $\Omega_{\rm open}$ is the angular frequency of the open
field lines (Contopoulos \& Spitkovsky 2006), $h$ is the thickness
of the gap, $h=1.1 \times 10^{4} \rho_{\rm 6}^{2/7} \Omega_{\rm
{open}}^{-3/7} B_{\rm {12}}^{-4/7} \, \rm cm$ and $\rho=2.3 \times
10^{8} R^{1/2}_{\rm 6} \Omega^{-1/2}_{\rm open}\,\rm cm$ is the
curvature radius of the open field lines ($\rho_{\rm
6}=\rho/10^{6} \rm cm$) (Ruderman \& Sutherland 1975). The
expression of $\eta$ can be rewritten as:
\begin{eqnarray}
\eta&=&\sin^{2}\alpha + 4.96\times 10^{2} \kappa \nonumber \\
&& \times (1-\frac{\Omega_{\rm death}}{\Omega})^{6/7}R_6^{-19/7}
B_{12}^{-8/7}\Omega^{-15/7}  \cos^{2}\alpha
\end{eqnarray}
The exponent of $(1-\frac{\Omega_{\rm death}}{\Omega})$ is the
minimum braking index in the VG(CR) model, $n=6/7$ (Li et al.
2014). In the wind braking model, the minimum braking index is
always around one (Li et al. 2014). Therefore, it is taken as one
in equation (\ref{eqn_Edotdeath}) (for all the acceleration
models). During the numerical calculations, the neutron star
radius is taken as $10^6 \,\rm cm$ (i.e., $R_6 =1$).

\section{Asymptotic behaviros in Figure 4 and 5}

In the pulsar wind model, the pulsar is braked down by the
combination of magnetic dipole radiation and particle wind. The
effect of particle wind is more and more important with pulsar
spinning down. As shown in Figure \ref{figparameters}, the period
and period derivative evolve up sharply with age. If the pulsar
braking is dominated by magnetic dipole radiation, equation
(\ref{nudotpowerlaw}) can be written as:
\begin{equation}
\dot{\nu}=-k\nu^{3}.
\end{equation}
the evolution of spin frequency is $\nu = (2kt)^{-1/2}$ (assuming
initial spin frequency is very large). Correspondingly, the period
and period derivative evolution with time are respectively: $P
=(2kt)^{1/2}$, and $\dot{P} = k(2kt)^{-1/2}$. If the pulsar is
braked down mainly by particle wind, the braking index is about
$1$. The spin-down power law is :
\begin{equation}
\dot{\nu}=-k\nu.
\end{equation}
the evolution of spin frequency is $\nu \propto \exp{(-kt)}$.
Correspondingly, the period and period derivative evolution with
time are respectively: $P \propto \exp{(kt)}$, and $\dot{P}
\propto \exp{(kt)}$. Clearly, the period and period derivative
evolve faster by the exponential form.

\label{lastpage}

\end{document}